\documentclass{aa}

\usepackage{txfonts}
\usepackage{graphicx}
\usepackage{natbib}
\bibpunct{(}{)}{;}{a}{}{,}

\newcommand{\zem}{\mbox{$z_{\rm em}$}}
\newcommand{\VmI}{\mbox{{$V$}{$-$}{$I$}}}
\newcommand{\BmV}{\mbox{{$B$}{$-$}{$V$}}}
\newcommand{\Oiii}{\mbox{{\rm O}\thinspace{\sc iii}}}
\newcommand{\Mgii}{\mbox{{\rm Mg}\thinspace{\sc ii}}}

\begin{document}


\title{Discovery of nine quasars behind the Large Magellanic
Cloud\thanks{Based on observations collected at the Magellan Baade
6.5-m telescope and at the European Southern Observatory, Chile (ESO
Programme 074.A-0069).}}

\author{A. Dobrzycki\inst{1}
\and
L. Eyer\inst{2}
\and
K. Z. Stanek\inst{3}\fnmsep\inst{4}
\and
L. M. Macri\inst{5}\fnmsep\thanks{Hubble Fellow}}

\offprints{A. Dobrzycki}

\institute{European Southern Observatory,
Karl-Schwarzschild-Strasse 2, D-85748 Garching bei M\"unchen, Germany\\
\email{adam.dobrzycki@eso.org}
\and
Observatoire de Gen\`eve, CH-1290 Sauverny,
Switzerland\\
\email{laurent.eyer@obs.unige.ch}
\and
Harvard-Smithsonian Center for Astrophysics, 60
Garden Street, Cambridge, MA 02138, USA\\
\email{kstanek@cfa.harvard.edu}
\and
Department of Astronomy, The Ohio State University, Columbus, OH 43210
\and
National Optical Astronomy Observatory, 950 North
Cherry Avenue, Tucson, AZ 85719, USA\\
\email{lmacri@noao.edu.}
}

\date{Received / Accepted }


\authorrunning{Dobrzycki et al.}


\abstract{

We present the discovery of nine quasars behind the Large Magellanic
Cloud, with emission redshifts ranging from 0.07 to 2.0. Six of them
were identified as part of the systematic variability-based search for
QSOs in the objects from the OGLE-II database. Combination of
variability-based selection of candidates with the candidates' colours
appears to be a powerful technique for identifying quasars,
potentially reaching $\sim$50\% efficiency. We report an
apparent correlation between variability magnitude and variability
timescale, which -- if confirmed -- could put even more constraints on
QSO candidate selection. The remaining three quasars were identified
via followup spectroscopy of optical counterparts to X-ray sources
found serendipitously by the {\em Chandra X-ray Observatory}
satellite. Even though the locations of the candidates were
quite uniformly distributed over the LMC bar, the confirmed QSOs
all appear near the bar's outskirts.


}

\maketitle


\section{Introduction\label{sec:introduction}}

Recent advances in observational techniques and the availability of
large variability monitoring databases have opened up a possibility
for searches for quasars behind dense stellar fields. Until recently,
those areas were avoided in QSO surveys, primarily because of
candidate confusion problems. At the same time, such quasars are very
interesting, e.g.\ as background sources for absorption studies (e.g.\
\citealp{proc2002}), as fixed reference points for dynamical studies
(e.g.\ \citealp{angu2000}), or even as means to determine very
accurate distances to the foreground objects \citep{drai2004}. The
launch of the {\em Chandra X-ray Observatory\/} -- with its superb
pointing accuracy and spatial resolution -- enabled X-ray based
searches, while microlensing campaigns (such as MACHO and OGLE)
provided the possibility for variability-based identifications.

The fields of particular interest are nearby galaxies, such as the
Magellanic Clouds. Application of new techniques resulted in the
number of known quasars in the general direction of the Clouds more
than trebling since 2002
\citep{dobr2002,dobr2003a,dobr2003b,geha2003}, and this difference is
more dramatic in the direction of the inner parts of the Clouds, where
almost all presently known QSOs were discovered in recent years.

Despite the remarkable success shown by these methods, there remained
an area which was still mostly unexplored: the bar of the
\object{Large Magellanic Cloud}, where only four quasars were
known. At the same time, this was the region where finding QSOs would
be most interesting for studying the dynamics of the LMC. The expected
proper motion at the distance to the LMC is of the order of only few
miliarcseconds per year (\citealp{angu2000} and references
therein). To study bulk motions it is thus critical to have a sizable
number of fixed reference points placed in ``strategic'' positions,
preferably behind the bar.

There were two main reasons for the scarcity of known QSOs in this
region. First, the X-ray-based searches -- which are responsible for
three out of four quasars behind the LMC bar -- are limited by the
relatively small number of {\em Chandra\/} pointings towards the LMC;
on top of that the known fields have been only partially
explored. Second, the only variability-based sample that has
so far been followed up spectroscopically is the MACHO sample
\citep{geha2003}. This sample -- although it is the biggest
contributor to the number of known quasars in the vicinity of the LMC
-- has not been followed up completely, and the completed part is
somewhat biased towards the outer parts of the LMC.

In this paper we present the results of the search for QSOs behind the
Large Magellanic Cloud. The variability-based identification of
candidates in the OGLE-II \citep{udal2000,zebr2001} data, which cover
4.6~deg$^2$ containing the bar of the LMC, was performed by
\citet{eyer2002}, and this list was the main part of our candidate
pool. We also followed up on several candidates from the X-ray
selected sample \citep{dobr2002}.

This work is a part of an ongoing project
\citep{dobr2002,dobr2003a,dobr2003b}. The variability-based technique
in the vicinity of the \object{Small Magellanic Cloud} turned out to
be very efficient, nearing 40\%, although the sample is rather small
(five quasars among seventeen candidates). Due to several
effects -- some related to the intrinsic differences between the
Clouds and some related to the nature of the OGLE data -- the list of
candidates in the direction of the LMC was expected to be more
contaminated with non-quasars. See Sections~2 and 3 in
\citet{eyer2002} for a discussion of those effects.

\section{New quasars\label{sec:newqsos}}

\subsection{Observations and identifications\label{subsec:observations}}

The spectroscopic observations of the candidates were performed on
several nights using two instruments. On 2002 January 22-23, 2002
November 07 and 17 we used the Magellan Baade 6.5-meter telescope, the
LDSS-2 imaging spectrograph with 2048$\times$2048 SITe\#1 CCD,
1.03~arcsec slit and 300~l/mm MedBlue grism, yielding a nominal
resolution of 13.3~\AA. On 2004 December 13-17 we used the ESO
3.6-meter telescope, the EFOSC2 imaging spectrograph and Loral/Lesser
2048$\times$2048 CCD\#40 detector, 1~arcsec slit and 300~l/mm grism
\#06, yielding a nominal resolution of 12.9~\AA.

During all nights, we observed \citet{hamu1992} spectrophotometric
standards and took exposures of He-Ne-Ar lamps for wavelength
calibrations. All spectra were reduced in the standard way using IRAF.


\begin{figure*}
\centering
\includegraphics[width=17cm]{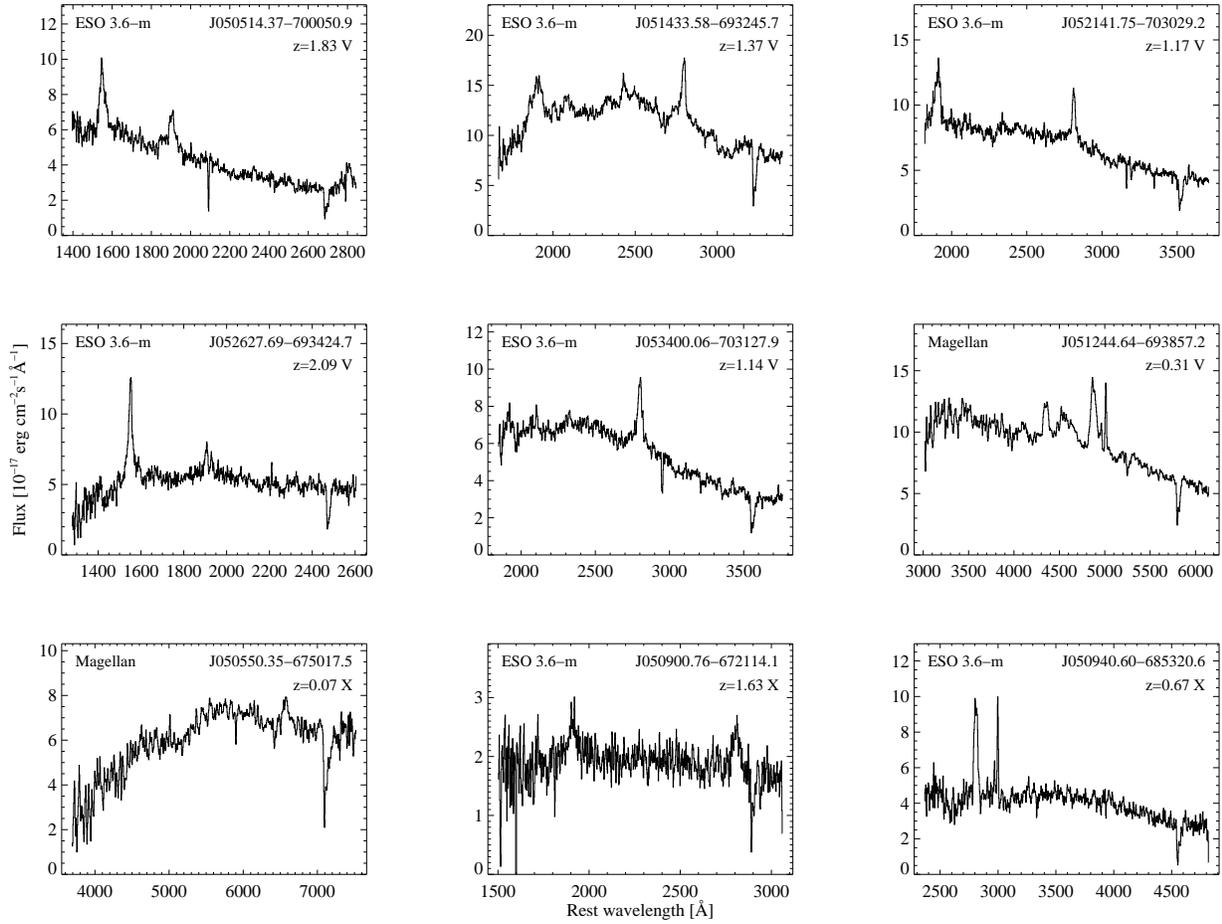}
\caption{Rest-frame discovery spectra of quasars presented in this
paper. First two rows show variability-selected QSOs, third row shows
X-ray-selected objects.\label{fig:spectra}}
\end{figure*}


\begin{figure*}
\centering
\includegraphics[width=17cm]{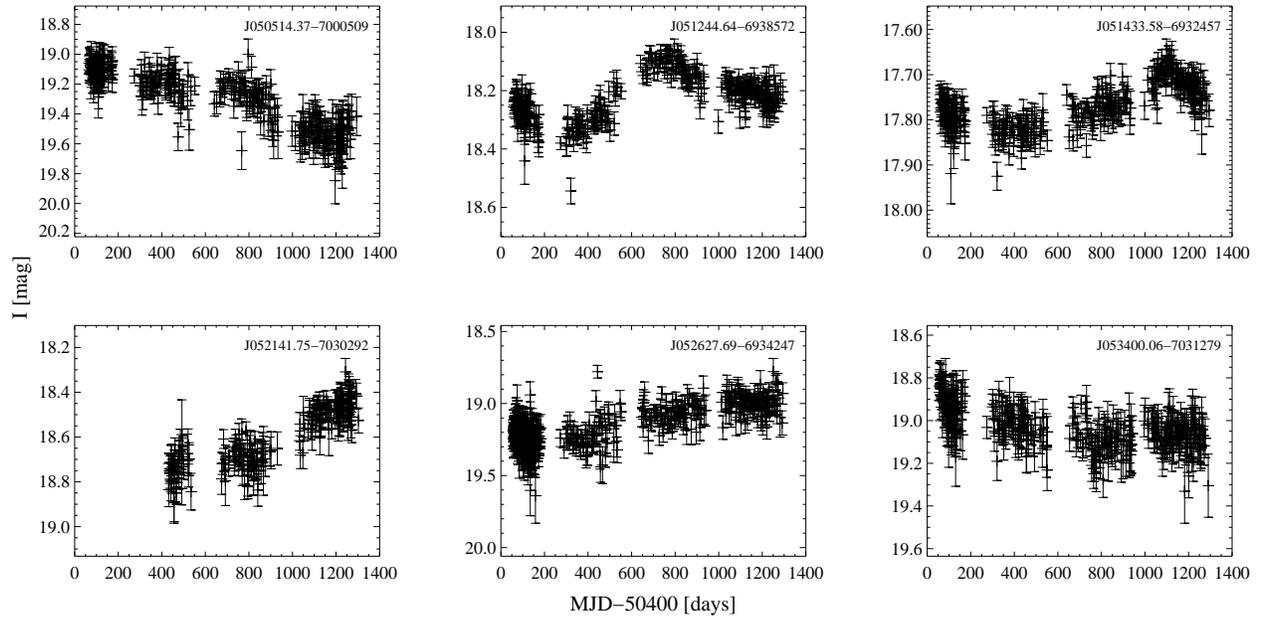}
\caption{OGLE \citep{zebr2001} light curves for six
variability-selected quasars (top six in
Figure~\ref{fig:spectra}). MJD 50,400 corresponds to UT 1996 November
13.\label{fig:lightcurves}}
\end{figure*}


\begin{table}
\begin{minipage}[t]{\columnwidth}
\caption{New quasars.}
\label{tab:quasars}
\centering
\renewcommand{\footnoterule}{}
\begin{tabular}{ccccc}
\hline\hline
J2000.0 coordinates & \zem & $V$ [mag] & Sel. & 
ID\footnote{Candidate identification in \citealp{eyer2002}} \\
\hline
\object{J050514.37--700050.9} & 1.83 & 20.08 & Var.  & L98  \\
\object{J051244.64--693857.2} & 0.31 & 19.51 & Var.  & L81  \\
\object{J051433.58--693245.7} & 1.37 & 18.87 & Var.  & L84  \\
\object{J052141.75--703029.2} & 1.17 & 19.31 & Var.  & L154 \\
\object{J052627.69--693424.7} & 2.09 & 19.81 & Var.  & L24  \\
\object{J053400.06--703127.9} & 1.14 & 19.39 & Var.  & L1   \\
\hline
\object{J050550.35--675017.5} & 0.07 & 19.00 & X-ray &      \\
\object{J050900.76--672114.1} & 1.63 & 20.58 & X-ray &      \\
\object{J050940.60--685320.6} & 0.67 & 19.79 & X-ray &      \\
\hline
\end{tabular}
\end{minipage}
\end{table}


The followup observations revealed nine previously unknown QSOs. Their
rest frame spectra are shown on Figure~\ref{fig:spectra} and a summary
of their properties is shown in Table~\ref{tab:quasars}.

\subsection{Variability selected quasars\label{subsec:varqsos}}

The set of candidates was constructed from the objects in
the OGLE-II database of variable objects \citep{zebr2001} using
several criteria. In addition to colour-magnitude and colour-colour
criteria, which removed many variable stars with known characteristics
(e.g. Cepheids), a more elaborate criterion, aimed at selecting
objects which (like QSOs) show larger variability on longer time
scales, was applied. This was done by calculating variogram, a
function showing time lag versus the spread of all magnitude
differences than the lag. The variogram is steeper for objects which
behave in the desired way, and an empirical cutoff on the variogram
slope for long timescales was applied. See \citet{eyer2002} for
details.

In total, the list of variability-selected QSO candidates contained
118 objects.  One of them, object L92 in \citet{eyer2002}, was
identified as a $\zem=0.61$ QSO in the X-ray-based search of
\citet{dobr2002}. Of remaining objects, six were also identified as
QSO candidates in the MACHO database by \citet{geha2003}. Two of those
objects, L114 and L155, were identified as quasars at, respectively,
$\zem=1.81$ and $\zem=0.90$; the other four objects turned out to be
stars.

During our observing runs, we completed observations of 108 objects
out of the remaining 111 candidates, identifying six of them as
QSOs. OGLE light curves of those six quasars are shown in
Figure~\ref{fig:lightcurves}. We do not see qualitative differences
between quasars and stars in the appearances of their light curves,
but see Section~\ref{subsec:variability} below for discussion of
possible quantitative difference between them in how variogram slope
relates to variability scale.

We consider the sample to be fully observed, since the remaining
three objects are very unlikely to be quasars, given their distinctly
non-QSO colour properties (see Section~\ref{subsec:colcol}).

Combined with three quasars identified previously, we found the
original variability-selected sample to include nine identified QSOs,
i.e.\ a $\sim$8\% success rate. This is lower that the rate seen in a
similar search in the direction of the Small Magellanic Cloud
\citep{dobr2003a} but, as mentioned earlier, not unexpected. The
variability-selected quasars have emission redshifts ranging from 0.31
to 2.09, and their discovery spectra (Fig.~\ref{fig:spectra}) seem to
show only typical quasar features.

\subsection{X-ray selected quasars\label{subsec:xrayqsos}}

During the runs we took several spectra of the X-ray selected
candidates from the list from \citet{dobr2002}. We refer the reader to
this paper for detailed description of the sample; we only mention
here that this sample is in no way homogeneous, therefore we will not
derive any quantitative properties regarding the efficiency of the
method or completeness of the resulting quasar harvest.

In total, we observed 30 objects. Of those, three turned out to be
quasars, at emission redshifts of 0.07, 0.67 and 1.63. We note here
that the first of them, QSO J050550.35--675017.5, is a relatively
strong and highly variable X-ray source (observed {\em Chandra} ACIS-7
count rate of 0.12-0.38 counts/sec), and perhaps could be used for
geometrical LMC distance determination using the method described in
\citet{drai2004}, although it is likely that a next generation X-ray
telescope would be needed. We also note that this object was detected
in the analysis of the XMM-Newton data by \citet{shty2005}, but not
identified as a QSO.

We also note that the narrow emission lines seen in the spectrum of
QSO J050940.60--685320.6 near $\lambda_{\rm rest}=3000$\AA\ are not
intrinsic to the quasar. The line of sight to this object intersects
an emission nebula and those features are spectral extraction
residuals from the [\Oiii] lines. The strong emission line is real;
lack of other emission features in the spectrum identifies it as
\Mgii$\lambda$2798 at $\zem=0.67$.

The analysis of the properties of these and other QSOs seen in X-rays
will be presented in the forthcoming paper. The non-quasars among the
X-ray-selected objects were mostly low mass X-ray binaries and hot
stars. We will present the analysis of those and other X-ray objects
observed during the course of our project in a separate paper.

\section{Properties of variability-selected quasars\label{sec:properties}}

Combining objects presented here with objects identified in previous
studies, we end up with fourteen quasars selected from the OGLE-II LMC
and SMC databases based on their variability: in the SMC
five QSOs identified by \citet{dobr2003a}, and in the LMC: six
identified in this paper, two identified by \citet{geha2003} and one
by \citep{dobr2002}. This sample is quite well defined: all those
objects were selected using the same technique from homogeneous, good
quality photometric data. We will present below the analysis of some
observable properties of those objects.

\subsection{Colour-colour diagram\label{subsec:colcol}}


\begin{figure}
\resizebox{\hsize}{!}{\includegraphics{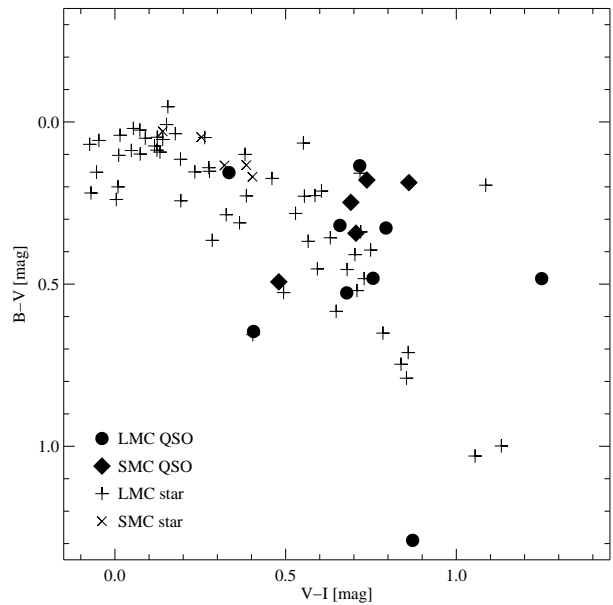}}
\caption{The \VmI\ vs.\ \BmV\ colour-colour diagram for objects in the
direction of the LMC and SMC. Filled symbols show QSOs. The
preferred location for quasars can be clearly seen.\label{fig:colcol}}
\end{figure}


In \citet{dobr2003a} we found that variability-selected quasars could
be distinguished from non-QSOs based on the \VmI\ colour, since
quasars were typically redder than other variable candidates. This
effect is clearly seen with new quasars added. In
Figure~\ref{fig:colcol} we show the \VmI\ vs.\ \BmV\ colour-colour
diagram for all LMC and SMC candidates. One can clearly see that
setting a cutoff at $\VmI \approx 0.3$ would dramatically increase the
efficiency of the variability-based selection method, to 40--50\%.

We note here that the three objects which were not observed -- L142,
L143 and L146 in \citet{eyer2002} -- all have $\VmI \lesssim
0$. Their location on Fig.~\ref{fig:colcol} indicates that they are
rather unlikely to be quasars.

It is, of course, a very well known fact that quasars
preferable occupy specific regions in colour-colour diagrams and this
information was applied to quasar surveys (e.g.\
\citealp{cris1989,newb1997,newb1999,rich2001}). However, we stress
that the diagram presented in Fig.~\ref{fig:colcol} only shows the
distinction between colour properties of stars and quasars which were
QSO candidates, already preselected with colour-based
criteria. Applicability of this plot is therefore limited to our
sample.

\subsection{Brightness-colour relation\label{subsec:slope}}

Several papers (e.g.\ \citealp{palt1994,helf2001,vand2004} and
references therein) indicate existence of a brightness-colour relation
in variable QSOs: quasars tend to be bluer in higher states.

In addition to a few hundred measurements in the $I$ band, the OGLE-II
database provides a few tens of measurements in the $B$ and $V$
bands. It is therefore possible to investigate this effect in our
sample.


\begin{figure}
\resizebox{\hsize}{!}{\includegraphics{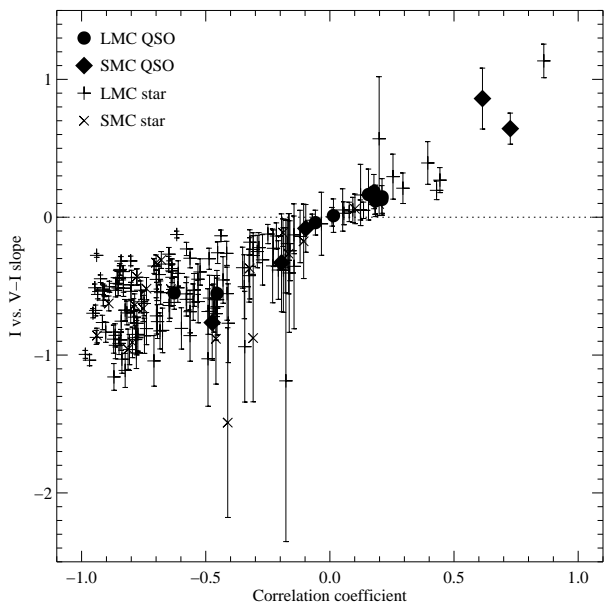}}
\caption{Parameters of the $I$ vs.\ \VmI\ relation: Pearson's
correlation coefficient versus slope for all identified objects in the
direction of LMC and SMC. Filled symbols show
quasars. See text for discussion.\label{fig:slope}}
\end{figure}


The diagram in Figure~\ref{fig:slope} shows the parameters of the $I$
vs.\ \VmI\ relation obtained from linear least square fits. The x-axis
shows the Pearson's correlation coefficient
while the y-axis shows the fitted slope and corresponding
uncertainty. The objects in which the ``brighter-bluer'' effect was
significant would occupy the upper right corner of
Fig.~\ref{fig:slope}, and their slopes would be significantly greater
than zero.

One can clearly see that this effect is not generally present in our
data. With the exception of two objects (both behind the SMC: S1 and
S12 in \citealp{eyer2002}), most of our quasars do not show this
effect and some even give an indication for opposite relation. This is
somewhat curious since the effect is rather well established in the
literature and one would expect that our set of objects was large
enough to show it.

We note that the behaviour of the \BmV\ colour is quantitatively
similar.

One cautionary note here is that sometimes the $BVI$ spectral bands
include QSO emission lines in our objects. This can weaken the effect,
since the lines may account for a sizeable fraction of the observed
flux, while the effect is associated with continuum emission. Possible
variability of emission lines -- e.g.\ the Baldwin effect -- may
affect the result in our case.

We also note that if this effect was indeed seen it could serve as an
auxiliary selection criterion since variable stars clearly do not
follow this relation.

\subsection{Amplitude-timescale relation\label{subsec:variability}}


\begin{figure}
\resizebox{\hsize}{!}{\includegraphics{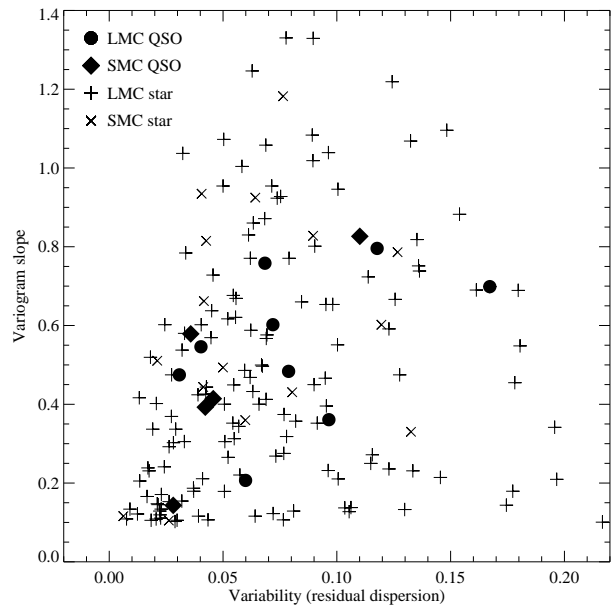}}
\caption{Variability versus variogram slope for identified
objects. Filled symbols show QSOs.\label{fig:variability}}
\end{figure}


Analysis of variability properties of the objects in the sample
appears to show an interesting phenomenon. In
Figure~\ref{fig:variability} we plot the residual variability
dispersion in OGLE-II data versus the variogram slope (see
\citealp{eyer2002}) for all objects with known identifications. 

There is no correlation between those two quantities for QSO
candidates that turned out to be stars. However, a decent correlation
($R=0.64$) is seen for quasars, indicating that the magnitude of the
QSO brightening is correlated with the timescale in which it
occurs. In other words, unlike stars, quasars do not show large
outbursts or long periods of relative calmness.

We emphasize that the above statements only apply to stars in the QSO
candidate sample and not to stars in general. A similar effect would
be seen in several types of variable stars, which were removed from
the original QSO candidate sample by magnitude and colour cuts.

This effect should be verified on a larger sample. If it is confirmed,
it can potentially be used to increase the efficiency of the candidate
selection, by filtering out those of the stars that lie above and
below the region in Figure~\ref{fig:variability} in which QSOs are
found.


\begin{figure*}
\centering
\includegraphics[width=17cm]{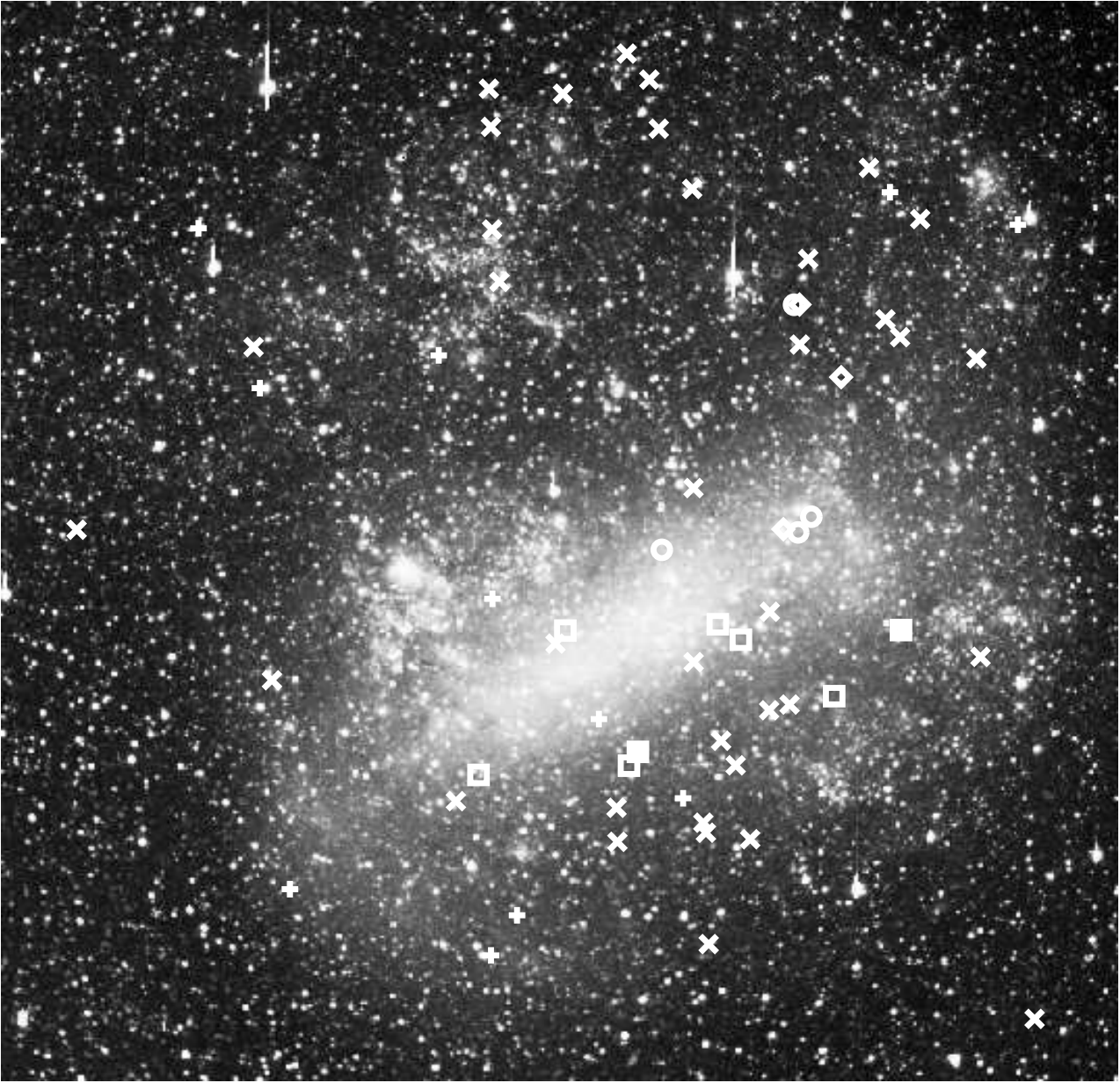}
\caption{Quasars in the vicinity of the Large Magellanic Cloud. The
image is roughly $7^\circ\times7^\circ$. Open squares and diamonds
show the positions of, respectively, variability- and X-ray-selected
quasars presented in this paper. The filled squares show
variability-selected QSOs from our candidate list confirmed by
\citet{geha2003}. The ``X'' marks show the variability-selected
quasars from \citet{geha2003}. Open circles show four X-ray selected
QSOs from \citet{dobr2002}. Crosses show the quasars known before
2002; a dramatic improvement in both the number of known objects and
the coverage is clearly seen. Only few QSOs are still known behind the
Cloud's bar. LMC image courtesy of G.~Bothun.\label{fig:lmcimage}}
\end{figure*}


\section{Discussion\label{sec:discussion}}

The quasars presented in this paper have increased the number of known
QSOs in the direction of LMC by 18\%. On Figure~\ref{fig:lmcimage} we
present the locations of the new quasars, together with the locations
of previously known QSOs. In total, this picture shows 61
quasars. The image shows a large qualitative jump in the number of
known objects since 2002.

However, the area in the direction of the LMC bar appears to
be underrepresented. This is somewhat surprising. The QSO candidates
were distributed quite uniformly over the bar, while the newly
discovered objects are all on its outskirts. It is possible that some
effects -- for example, increased extinction or source crowding
affecting variability analysis -- do influence the candidate selection
in the densest fields. However, since the number of object is still
rather small, we are not in a position to determine whether a true
bias is involved.

We confirm the result seen in the SMC data \citep{dobr2003a} that the
addition of the colour criterion to the variability selection
procedure will lead to much higher efficiency in quasar searches,
since QSOs are typically redder than other variability-selected
objects. As expected, this distinction for the LMC objects was
somewhat less strict compared to the SMC, but even here it would lead
to very high, approaching 50\%, success rates.

In our data we see a marginally significant correlation between the
variability magnitude and the variability timescale. This correlation
is entirely absent for variable stars from the QSO candidate pool. If
this effect is confirmed, it can be used to increase the method
efficiency even more.

Quasars from our sample do not generally follow the ``brighter-bluer''
relation seen in larger samples, such as SDSS. However, this may be
affected by variability of the QSO emission lines, which enter
observed photometry bands.

We plan to utilise the variability selection method for searching for
quasars behind the Galactic Bulge, in low extinction windows monitored
by OGLE-II (see \citealp{sumi2005}). A successful search would carry
great hope for studies of the dynamics of the Bulge, since the
expected proper motions are much larger than in the Magellanic Clouds.

\begin{acknowledgements}

We thank B. Czerny, B. Paczy\'nski, S. Paltani, G. Richards and
A. Siemiginowska for helpful discussions, and the referee, M. Geha,
for thorough and helpful review. We thank I. Saviane and the La Silla
personnel for excellent support during our observing run at the La
Silla observatory, and P. Challis for observing some candidates for us
with the Magellan telescope.  This research has made use of the
NASA/IPAC Extragalactic Database (NED), operated by the Jet Propulsion
Laboratory, California Institute of Technology, under contract with
NASA, and of the SIMBAD database, operated at CDS, Strasbourg,
France. LMM was supported by the Hubble Fellowship grant HF-01153.01-A
from the Space Telescope Science Institute, which is operated by the
Association of Universities for Research in Astronomy, Inc., under
NASA contract NAS5-26555.

\end{acknowledgements}



\end{document}